\pdfoutput=1
\documentclass{ieeeaccess}

\usepackage{cite}
\usepackage{amsmath,amssymb,amsfonts}
\usepackage[scaled]{beramono}
\usepackage{booktabs}
\usepackage{enumitem}
\usepackage{graphicx}
\usepackage{textcomp}
\usepackage{microtype}
\usepackage{needspace}
\usepackage[group-separator={,}]{siunitx}
\usepackage[caption=false,subrefformat=parens]{subfig}
\usepackage{listings}
\usepackage{url}
\usepackage{xcolor}
\definecolor{accessblue}{cmyk}{1,0.3,0,0.2}
\AtBeginDocument{%
  \DeclareFontShape{T1}{ptm}{n}{n}{<->ssub*ptm/m/n}{}%
  \DeclareFontShape{T1}{ptm}{n}{it}{<->ssub*ptm/m/it}{}%
  \DeclareFontShape{T1}{ptm}{n}{sl}{<->ssub*ptm/m/sl}{}%
  \DeclareFontShape{T1}{ptm}{n}{sc}{<->ssub*ptm/m/sc}{}%
  \bodyfont%
}

\makeatletter
\def\if@biographyTOCentrynotmade{\if@IEEEbiographyTOCentrynotmade}
\let\@biographyTOCentrynotmadefalse\@IEEEbiographyTOCentrynotmadefalse
\let\c@biography\c@IEEEbiography
\let\p@biography\p@IEEEbiography

\makeatother

\providecommand{\xfigwd}{0pt}

\graphicspath{{figs/}}

\definecolor{lstkw}{HTML}{0057B7}
\definecolor{lstcmt}{HTML}{6A737D}
\definecolor{lststr}{HTML}{067D17}
\lstdefinestyle{paper}{
  basicstyle=\ttfamily\small,
  keywordstyle=\color{lstkw}\bfseries,
  commentstyle=\color{lstcmt}\itshape,
  stringstyle=\color{lststr},
  numbers=left, numberstyle=\tiny\color{gray}, numbersep=8pt,
  frame=lines, framesep=6pt,
  backgroundcolor=\color{gray!5},
  breaklines=true, showstringspaces=false, tabsize=2,
}

\begin{document}

\history{Date of publication xxxx 00, 0000, date of current version xxxx 00, 0000.}
\doi{10.1109/ACCESS.XXXX.DOI}

\title{Performance Evaluation of RED-ONION:\\A High-Speed Disk-to-Disk Transfer System}

\author{
  \uppercase{Keichi Takahashi}\authorrefmark{1},
  \IEEEmembership{Member, IEEE},
  \uppercase{Hiroaki Kataoka}\authorrefmark{2},
  \uppercase{Takeo Hosomi}\authorrefmark{2,1},\\
  \uppercase{Ayahiro Takaki}\authorrefmark{3},
  \uppercase{Yasunori Kakizawa}\authorrefmark{3},
  \uppercase{Shuichi Ihara}\authorrefmark{4},
  \uppercase{Nobuaki Hashizume}\authorrefmark{4},\\
  and \uppercase{Susumu Date}\authorrefmark{1},
  \IEEEmembership{Member, IEEE}}
\address[1]{D3 Center, The University of Osaka, Ibaraki, Osaka 567-0047, Japan
  (e-mail: takahashi.d3c@osaka-u.ac.jp; date@cmc.osaka-u.ac.jp)}
\address[2]{NEC Corporation, Minato-ku, Tokyo 108-8001, Japan
  (e-mail: hkataoka@nec.com; takeo.hosomi@nec.com)}
\address[3]{CLEALINK TECHNOLOGY Co., Ltd., Seika, Kyoto 619-0237, Japan
  (e-mail: takaki@clealink.jp; kakizawa@clealink.jp)}
\address[4]{DataDirect Networks Japan, Inc., Chiyoda-ku, Tokyo 102-0081, Japan
  (e-mail: sihara@ddn.com; nhashizume@ddn.com)}

\markboth
{Takahashi \headeretal: Performance Evaluation of RED-ONION}
{Takahashi \headeretal: Performance Evaluation of RED-ONION}

\corresp{Corresponding author: Keichi Takahashi (e-mail: takahashi.d3c@osaka-u.ac.jp).}

\begin{abstract}
Modern experimental instruments produce data faster than general-purpose file transfer interfaces
can move it, so delivery to the computing infrastructure has become a bottleneck in the research
process. At many universities and research institutes, moreover, the instruments that generate
research data and the high-performance computing systems that analyze it are separated both
geographically and organizationally, because each demands its own expertise and installation
environment. Connecting the two seamlessly is a pressing challenge for data-driven science. This
article presents RED-ONION, a high-speed disk-to-disk transfer system that connects research
facilities, on campus and beyond, to a computing center. The system combines data transfer nodes, a
dedicated high-bandwidth network, an all-flash parallel file system, and multi-threaded transfer
software that parallelizes network transmission and storage access. The design targets the wire
rate both along the entire path, from the read on the sender storage to the write on the receiver
storage, and for a single file between one pair of nodes rather than only in aggregate over many
files or nodes. We describe the end-to-end optimizations across the transfer software, the
operating system, and the storage that this requires. We evaluate a prototype deployed over a
\SI{100}{Gbps} transpacific path between Atlanta and Tokyo with a \SI{150}{ms} round-trip time, on
which a single \SI{1}{TB} file transfer reached \SI{90}{Gbps}, delivering a terabyte in
approximately \SI{95}{s}. Moving a dataset of this size therefore becomes a routine step, and the
computing center serves an instrument as if the two were co-located.
\end{abstract}


\titlepgskip=-15pt

\maketitle

\section{Introduction}\label{sec:introduction}
Scientific research increasingly relies on high-performance computing infrastructure to analyze the
real-world data produced by large-scale experimental instruments and to train machine learning
models on it, a trend referred to as \emph{AI for Science}~\cite{Wang2023}.
Modern experimental instruments generate data in unprecedented volumes and at unprecedented
rates. DNA sequencers~\cite{Schmidt2017}, radio telescopes~\cite{Scaife2020}, cryo-electron
microscopes~\cite{Baldwin2018}, and light sources~\cite{Sobolev2024,Obara2025} each produce on the
order of terabytes to petabytes of data per day.

\begin{figure*}
\centering
\includegraphics{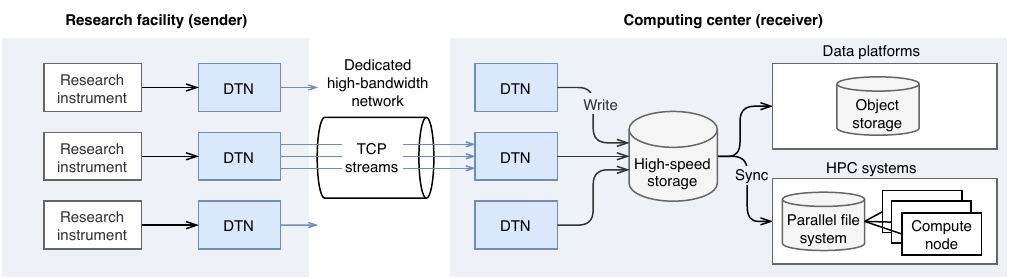}
\caption{Overview of the RED-ONION architecture.}
\label{fig:architecture}
\end{figure*}

An experimental facility and a computing center require different kinds of expertise to build and
operate. They also impose different requirements on the physical infrastructure, such as power,
cooling, and floor space. Universities and research institutes thus have no choice but to keep the
two apart, at different locations and under separate organizations. Research data must nonetheless
reach the computing center quickly for data analysis, large-scale simulation, and machine learning.
Connecting the two seamlessly is therefore a pressing challenge in accelerating AI for Science.

To aggregate and accumulate the research data produced across the campus of the University of
Osaka, our computing center built a data platform called the \underline{O}saka University
\underline{N}ext-generation \underline{I}nfrastructure for \underline{O}pen Research and
Innovatio\underline{N}~(ONION)~\cite{Date2023,Tanushi2024}, which unifies three components hosted
at the computing center for aggregating, storing, and publishing diverse research data. These
components are the parallel file system of the supercomputers, an S3-compatible object storage, and
a web-based file sharing service. Researchers have conventionally uploaded their data to ONION over
the Internet and the campus network through interfaces such as SCP/SFTP, the S3-compatible API, or
the web UI. None of these interfaces was designed for bulk data movement, and none of them reaches
a throughput that matches the bandwidth of the underlying network. As data volumes grow, the time
required to transfer data from research facilities on campus to ONION has therefore become a
bottleneck in the research process.

To shorten this transfer time, we are currently building a dedicated high-bandwidth fiber network
between the major research institutes on campus and the computing center, and placing hardware and
software optimized for data transfer at both ends of the network. This system, which we call the
\underline{R}esearch-\underline{E}nhance\underline{D} ONION~(RED-ONION), will enable high-speed
transfer of large volumes of data from research facilities to the computing center. The data
received there is then either stored in a data platform such as ONION for archiving and publishing,
or transferred onto the parallel file system of the supercomputers to be used for machine learning
and data analysis. The network described so far is confined to the campus, but many of the advanced
instruments on which researchers at the university depend, such as light sources and telescopes, are
operated at off-campus facilities. The same design applies to those facilities once the dedicated
network is extended over a wide-area research network.

In this article, we present the design of RED-ONION and describe the performance optimizations
required to transfer a single large file, as well as a bulk transfer of small files, from disk to
disk at the \emph{wire rate}. By wire rate, we mean a throughput close to the theoretical bandwidth
of the link. We then report performance evaluation results obtained on a prototype deployment of
RED-ONION, built to validate the design on the most demanding path it must support, a transpacific
path between a site in the United States and a site in Japan.

The contributions of this article are as follows:
\begin{itemize}
  \item We present the design of RED-ONION, an end-to-end disk-to-disk data transfer
    system built from data transfer nodes (DTNs), a dedicated high-bandwidth network, an all-flash
    parallel file system, and high-speed data transfer software.
  \item We describe the end-to-end performance optimizations, spanning the data transfer software,
    the operating system, and the storage system, that are required to achieve wire-rate transfer
    of a single large file between two DTNs.
  \item We report a performance evaluation of a prototype deployment over a transpacific network
    connecting Atlanta and Tokyo, demonstrating disk-to-disk transfer of a single \SI{1}{TB} file
    at \SI{90}{Gbps}.
\end{itemize}

The remainder of this article is organized as follows. Section~\ref{sec:challenges} describes the
challenges that RED-ONION addresses. Section~\ref{sec:mechanism} presents the design of the
high-speed disk-to-disk transfer system and its performance tuning. Section~\ref{sec:evaluation}
reports the performance evaluation on a prototype deployment. Section~\ref{sec:related} reviews
related work, and Section~\ref{sec:conclusion} concludes the article.

\section{Technical Challenges}\label{sec:challenges}
Data on its way from one disk to another passes through the sender's file system, page cache,
network stack, and network interface, then traverses the wide-area path and retraces the same
sequence in reverse. Each stage can throttle the transfer independently, and tuning any one of them
in isolation merely relocates the bottleneck. Closing the gap between the bandwidth of the link and
the disk-to-disk throughput a user observes therefore requires attention to the whole path, and
this tuning must not sacrifice the availability expected of a production service. RED-ONION
addresses the following three challenges:
\begin{enumerate}[label={\textbf{C\arabic*)}}, ref={\textbf{C\arabic*}}]
  \item\label{ch:endtoend} \textbf{End-to-end transfer performance optimization:} RED-ONION aims to
    attain wire-rate throughput along the entire disk-to-disk path, from reading data on the
    sender-side storage to writing it on the receiver-side storage. Wire-rate throughput must
    therefore be sustained not only over the network but also when reading from and writing to
    storage.
  \item\label{ch:onefile} \textbf{High-speed single-file transfer:} Many prior studies have reported
    achieving wire-rate aggregate throughput by transferring a large number of files
    concurrently or by transferring data in parallel across many
    DTNs~\cite{Allcock2005,Kettimuthu2018,Bird2011,Scaife2020}. In contrast, RED-ONION aims
    to achieve the wire rate when transferring one file between one pair of DTNs. A single file
    transfer must therefore fill the link, instead of reaching the wire rate only in aggregate over
    many files or many DTNs.
  \item\label{ch:operation} \textbf{Availability and scalability for production operation:}
    RED-ONION is not a research testbed, but is currently being deployed at the University of Osaka
    as part of the research data infrastructure. We also plan to extend it to other research
    institutions. It must therefore remain available in the event of DTN or disk failures and scale
    to additional sites, on and off campus, as demand grows.
\end{enumerate}

\section{High-Speed Disk-to-Disk Transfer System}\label{sec:mechanism}

\subsection{Overview}\label{sec:overview}

Figure~\ref{fig:architecture} illustrates the high-level architecture of RED-ONION. On the sender
side, a DTN reads a file from local high-speed storage such as NVMe SSDs and splits it
into many blocks, which are streamed across multiple parallel TCP connections over a dedicated
high-bandwidth network. On the receiver side, a peer DTN reassembles the blocks and
writes them to shared high-speed storage, from which the data is later ingested into archival
storage or the parallel file system of a supercomputer. Sustaining the wire-rate throughput
along this disk-to-disk path requires the following four elements to be jointly designed and tuned.
\begin{itemize}
  \item \textbf{Data transfer nodes:} A group of servers responsible for inter-site data transfer.
    Each DTN is equipped with a high-bandwidth network interface and connects to the high-bandwidth
    network. It also mounts the high-speed storage. The DTNs run the data transfer software and
    exchange data with the peer site.
  \item \textbf{High-bandwidth network:} A network that connects the sites using Ethernet at
    \SI{100}{Gbps} or higher. It uses dedicated fiber within a campus, and dedicated paths on a
    wide-area research network between distant sites.
  \item \textbf{High-speed storage:} An all-flash parallel file system mounted by
    the DTNs on the receiver side. It stores the data received by the DTNs and,
    as needed, moves the data to the object storage or parallel file system of the data platform.
  \item \textbf{Data transfer software:} A program that runs on the DTNs and
    performs the actual transfer. It splits a file into many blocks, distributes them across
    multiple connections, and performs storage reads and writes together with network send and
    receive at a high degree of parallelism.
\end{itemize}

The following subsections describe these elements in detail.

\subsection{Data Transfer Node and High-Bandwidth Network}
Each DTN is a server equipped with a multi-core CPU, a large amount of memory, high-bandwidth
network interface cards, and fast local or shared storage. The core count bounds the number of
TCP streams a node can drive, because the data transfer software dedicates one thread to each TCP
stream, and the DTN must drive as many TCP streams as are needed to fill the link with a single file
(\ref{ch:onefile}). The core counts of mainstream server CPUs already meet this requirement.
A receiver-side DTN stores data on a shared parallel file system, and thus carries storage traffic
in addition to transfer traffic. We therefore equip such a DTN with separate network interfaces so
that the two do not contend for the same link.

The DTNs at a site mount the same shared storage, so service can fail over to another DTN when one
of them fails, and the transfer capacity of a site grows by adding DTNs (\ref{ch:operation}). Since
a single DTN is designed to fill the link on its own (\ref{ch:onefile}), losing one of them does not
reduce the throughput of an individual transfer, and only the number of transfers that the site can
serve concurrently is affected.

The high-bandwidth network connects the sites using Ethernet at \SI{100}{Gbps} or higher, with jumbo
frames enabled end to end. Jumbo frames reduce the number of packets needed to sustain a
given throughput, and hence the per-packet processing overhead on the hosts.

\subsection{High-Speed Storage}
The high-speed storage on the receiver side is DDN EXAScaler, an all-flash parallel file system
based on Lustre~\cite{Schwan2003}, which the DTNs mount through a dedicated storage network. A file
must be written to this storage at the wire rate of the high-bandwidth network (\ref{ch:endtoend},
\ref{ch:onefile}), and this rate is usually beyond what a single storage device or server can
sustain. A parallel file
system meets the requirement by spreading the data of one file over many devices and servers.

\begin{figure}
\centering
\includegraphics{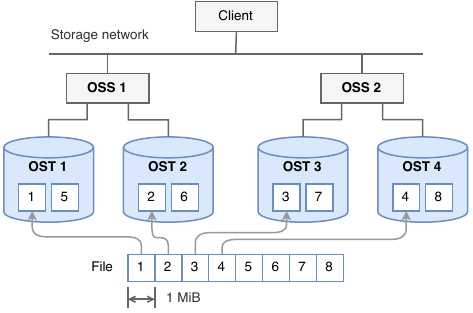}
\caption{Striping of a file across the OSTs of an EXAScaler file system, with a stripe count of four
and a stripe size of \SI{1}{MiB}.}
\label{fig:lustre}
\end{figure}

Figure~\ref{fig:lustre} shows how EXAScaler spreads file data in this way. The data is served by a
set of \emph{object storage servers} (OSSs), and each OSS owns one or more \emph{object storage
targets} (OSTs), the block devices on which the data actually resides. The content of a file is
distributed over one or more OSTs in fixed-size chunks called \emph{stripes}, so that the I/O
requests of a single file are served by multiple OSSs and their OSTs in parallel. Two parameters
control this layout. The \emph{stripe count} is the number of OSTs a file is spread over, and the
\emph{stripe size} is the number of consecutive bytes written to one OST before EXAScaler moves on
to the next.

The storage must also keep serving transfers even when hardware fails (\ref{ch:operation}). We
therefore build each OST as a RAID volume, which absorbs the failure of an individual drive, and
deploy the OSSs in high-availability pairs attached to the same drives, so that the OSTs of a failed
OSS fail over to its partner OSS.

\begin{figure}
\centering
\includegraphics{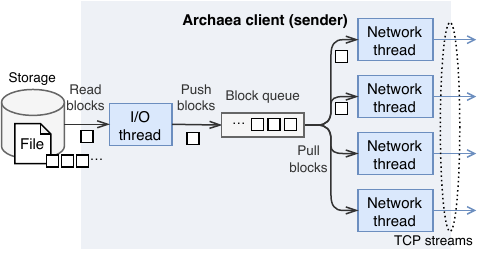}
\caption{Internal design of the Archaea client.}
\label{fig:hcp}
\end{figure}

\subsection{Data Transfer Software}
RED-ONION performs the actual data movement with \emph{Archaea},
a high-throughput file transfer suite developed by CLEALINK
TECHNOLOGY.\footnote{\url{https://hcp.clealink.jp/}} The Archaea client transfers data to and from a
remote Archaea server and supports several transports, among them TCP, a UDP-based protocol, and
WebSocket. RED-ONION uses TCP, because the network it runs over is dedicated to data transfer and
carries no competing traffic, so the congestion and packet loss that
motivate a UDP-based transport do not arise. Rather than streaming a file as a single byte stream,
the Archaea client aggregates file data into fixed-size \emph{content blocks} and distributes them
across multiple TCP connections, transferring them in parallel so as to fill links whose bandwidth
exceeds the throughput of a single connection (\ref{ch:onefile}).\footnote{\url{https://support.bytix.tech/docs/archaea/tools/1.5/A_overview/A05_function_transfer/}}

Figure~\ref{fig:hcp} shows the internal design of the Archaea client, which is organized as a
multi-threaded pipeline that decouples storage access from network transmission. Reading and
transmitting therefore proceed at the same time, and neither the storage nor the network sits idle
while the other works (\ref{ch:endtoend}). A single I/O thread reads content blocks from storage
and places them into an in-memory queue, and a group of network threads each pull blocks from this
queue and transmit them, with one thread driving one TCP stream. The number of streams scales with
the available CPU cores. Because the network threads pull blocks from a shared queue, the load is
balanced across connections and a temporarily slow connection does not stall the others. The I/O
and network paths operate on blocks of the same size, so a block read from storage is handed to a
network thread without additional copies.

\begin{figure}
\centering
\includegraphics{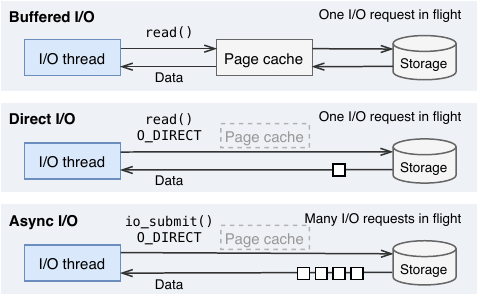}
\caption{Comparison of the three I/O modes supported by Archaea, shown for a read.}
\label{fig:io-modes}
\end{figure}

To prevent the storage from becoming the bottleneck at these rates, Archaea supports three I/O
modes.\footnote{\url{https://support.bytix.tech/docs/archaea/tools/1.5/D_commandRef/D01_hcp/}}
Figure~\ref{fig:io-modes} compares how data moves between storage and user space in each mode.
\emph{Buffered I/O} reads and writes through the page cache. \emph{Direct I/O} bypasses the page
cache and moves data directly between storage and user space. \emph{Asynchronous I/O} combines
direct I/O with the Linux native asynchronous I/O interface, accessed through the asynchronous I/O
system calls,\footnote{\url{https://www.man7.org/linux/man-pages/man2/io_submit.2.html}} to keep
many I/O operations in flight and hide I/O latency.

\begin{figure*}
\centering
\includegraphics{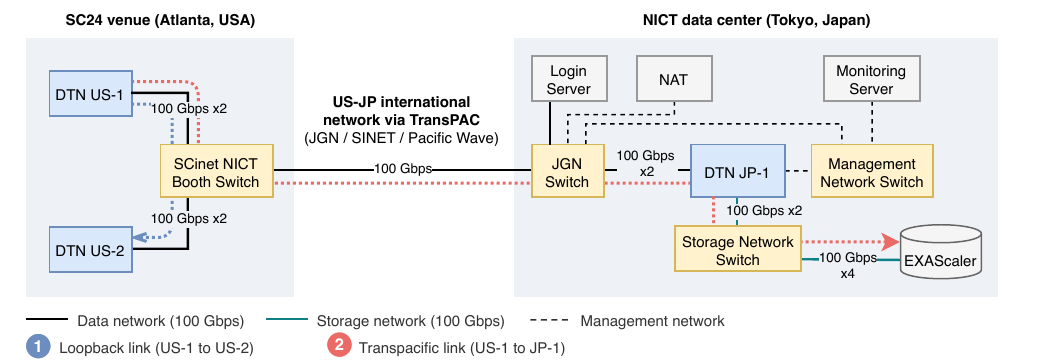}
\caption{Experimental environment at SC24.}
\label{fig:network}
\end{figure*}

\subsection{Performance Tuning}\label{sec:tuning}
Reaching the wire rate requires jointly tuning the data transfer software, the operating system, and
the storage. This subsection describes the rationale behind each setting, and the concrete values we
used are reported in Section~\ref{sec:setup}.

\subsubsection{Data Transfer Software}
The default Archaea client configuration is tuned conservatively for general-purpose use, so several
of its limits must be relaxed to exploit the full bandwidth of the network and storage. We use the
asynchronous I/O mode to bypass the page cache and overlap I/O operations, and encryption for
transfers within the trusted network is disabled to remove the associated overhead. We also enlarge
the transfer buffers that sit between the I/O thread and the network threads. These buffers absorb
the momentary difference in rate between the storage and the network, so that a period in which the
storage reads more slowly than the network transmits is covered from memory and the transfer rate is
sustained. Finally, we increase the content block size to amortize the per-block overhead of the
pipeline.

\subsubsection{Operating System}
On a long-distance high-bandwidth link, the sender must wait a round-trip time before an ACK packet
returns, so the data in flight is limited to the TCP window. When the window is smaller than the
\emph{bandwidth-delay product (BDP)}, the sender stalls waiting for ACK packets and cannot fill the
link (i.e., the long fat pipe problem~\cite{Jacobson1992}). On Linux, the socket buffers are
autotuned within the bounds given by \texttt{net.ipv4.tcp\_rmem} and \texttt{net.ipv4.tcp\_wmem},
which specify the minimum, initial, and maximum buffer size in bytes for the receive and send
buffers, respectively. By default the maximum is only a few mebibytes. This is sufficient within a
campus, where the round-trip time is under a millisecond, but it is significantly smaller than the
BDP of an intercontinental \SI{100}{Gbps} path. We therefore leave the minimum and initial values
unchanged and raise the maximum of both parameters.

\subsubsection{Storage}
By default the stripe count of files on EXAScaler is one, so a file resides entirely on a single
OST. This default suits HPC systems, where the file system is shared by many users and many files
are simultaneously accessed. Even with a stripe count of one, these files are stored on different
OSTs and the load is spread across the storage. RED-ONION instead aims to sustain the wire rate
while transferring a single file (\ref{ch:onefile}), and a stripe count of one confines that file to
one OST. We therefore set the stripe count to the total number of OSTs, so that the I/O of a single
file is spread across all OSTs.

The stripe size is chosen so that its product with the stripe count divides the size of an
individual I/O request, which for Archaea is the content block size. Striping turns each I/O request
into multiple remote procedure calls (RPCs) that are issued at the same time. When the I/O request
size is a multiple of the stripe size times the stripe count, the resulting RPCs reach every OST and
no OST sits idle.

\needspace{4\baselineskip}
\section{Performance Evaluation}\label{sec:evaluation}

\subsection{Experimental Setup}\label{sec:setup}
To demonstrate the feasibility of the RED-ONION concept, we designed a small-scale prototype
spanning two sites in the United States and Japan, and deployed and evaluated it during SC24, the
International Conference for High Performance Computing, Networking, Storage, and
Analysis,\footnote{\url{https://supercomputing.org/}} held at
the Georgia World Congress Center in Atlanta from November~17 to November~22, 2024.
Figure~\ref{fig:network} shows an overview of the prototype. Two DTNs were installed at the SC24
venue, and one DTN, together with the high-speed storage, was installed
at a data center of the National Institute of Information and Communications Technology
(NICT) in Tokyo. A monitoring server was set up at the same data center to collect and visualize the
network throughput as well as the CPU and memory utilization of each DTN.

\subsubsection{Data Transfer Nodes}
Table~\ref{tab:dtn} summarizes the configuration of the three DTNs. The two DTNs deployed in Atlanta
(DTN~US-1 and DTN~US-2) are each equipped with ten Micron 9100 PRO NVMe U.2 SSDs and a dual-port
Mellanox ConnectX-5 \SI{100}{Gbps} Ethernet adapter. The DTN deployed in Tokyo (DTN~JP-1) is
equipped with two dual-port Mellanox ConnectX-6 adapters, one used for EXAScaler traffic and the
other for data transfer. All DTNs run Rocky Linux~9.4 and have jumbo frames
(MTU \num{8972}) enabled.

\begin{table*}
  \centering
  \caption{Specifications of the data transfer nodes used in the evaluation.}
  \label{tab:dtn}
  \footnotesize
  \begin{tabular}{llll}
    \toprule
     & DTN~US-1 & DTN~US-2 & DTN~JP-1 \\
    \midrule
    Site & Atlanta venue & Atlanta venue & NICT (Tokyo) \\
    CPU & AMD EPYC 7302 & AMD EPYC 7402
       & 2$\times$ Intel Xeon Gold 6338 \\
       & (\SI{3.0}{GHz}, 16 cores) & (\SI{2.8}{GHz}, 24 cores) & (\SI{2.0}{GHz}, 32 cores each) \\
    Memory & \SI{256}{GiB} & \SI{256}{GiB} & \SI{512}{GiB} \\
    Data storage & 10$\times$ Micron 9100 PRO & 10$\times$ Micron 9100 PRO & Shared DDN EXAScaler \\
       & NVMe U.2 SSD (\texttt{mdadm} & NVMe U.2 SSD (\texttt{mdadm} &                    \\
       & RAID~0, XFS) & RAID~0, XFS) & \\
    Network & Dual-port Mellanox ConnectX-5 & Dual-port Mellanox ConnectX-5
       & 2$\times$ dual-port Mellanox ConnectX-6 \\
       & (\SI{100}{Gbps}) & (\SI{100}{Gbps}) & (\SI{100}{Gbps}) \\
    OS & Rocky Linux~9.4 & Rocky Linux~9.4 & Rocky Linux~9.4 \\
    \bottomrule
  \end{tabular}
\end{table*}

\subsubsection{Storage}
The shared high-speed storage is a DDN ES400NVX all-flash appliance populated with twenty
\SI{1.92}{TB} NVMe SSDs, providing an effective capacity of about \SI{30}{TB} and an aggregate
read/write performance of \SI{35}{GB/s}, connected over four \SI{100}{Gbps} Ethernet links. The
appliance exports an EXAScaler parallel file system (based on Lustre~2.14) over four OSSs and eight
OSTs, which DTN~JP-1 mounts through a switch. DTN~US-1 and US-2 each combine their ten NVMe SSDs
into a single \texttt{mdadm} RAID~0 volume formatted with the XFS file system.

\subsubsection{Network Paths}
We evaluated two transfer paths, each reflecting a different deployment scenario. One, which we call
the loopback path, represents the ingestion of data produced by a research instrument located on the
same campus. The other, the transpacific path, represents the ingestion of data produced by a
research facility located overseas.
\begin{itemize}
  \item \textbf{Loopback path.} DTN~US-1 transfers to DTN~US-2. Both nodes are in Atlanta, both read
    and write local NVMe, and both connect to the same switch in the SC24 booth, so the traffic
    never leaves the booth. This isolates the transfer from the instability of the wide-area
    network. This path is shown in blue and labeled~1 in Fig.~\ref{fig:network}.
  \item \textbf{Transpacific path (US--JP).} DTN~US-1 in Atlanta (local NVMe) transfers to
    DTN~JP-1 in Tokyo (shared EXAScaler) over a \SI{100}{Gbps} international link provided by NICT
    and routed via TransPAC\footnote{\url{https://www.transpac.org/}}
    (JGN\footnote{\url{https://www.jgn.nict.go.jp/english/}},
    SINET\footnote{\url{https://www.sinet.ad.jp/}}, and
    Pacific Wave\footnote{\url{https://pacificwave.net/}} on the Japan-side and exchange segments).
    This path is shown in red and labeled~2 in Fig.~\ref{fig:network}.
    The round-trip time between DTN~US-1 and DTN~JP-1 observed with \texttt{ping} was approximately
    \SI{150}{ms}.
\end{itemize}

\subsubsection{Monitoring}
The monitoring server located in Tokyo runs Grafana\footnote{\url{https://grafana.com/}} backed by
Prometheus,\footnote{\url{https://prometheus.io/}} and every DTN runs the Prometheus node exporter,
which exposes the CPU, memory, and network interface counters that Prometheus scrapes. Prometheus
samples these counters at most once per minute, which is too coarse to resolve how the throughput of
a transfer evolves over its lifetime. We therefore additionally installed
vnStat\footnote{\url{https://humdi.net/vnstat/}} on each DTN and recorded the per-interface transfer
volume every two seconds.

\subsubsection{Tuning Parameters}
We applied the tuning described in Section~\ref{sec:tuning} with the following values. In the
Archaea client, we enlarged the total transfer buffer from \SI{4}{GB} to \SI{220}{GB} and the
per-connection buffer from \SI{100}{MB} to \SI{300}{MB}, and set the content block size to
\SI{64}{MiB}. For the socket buffers, the BDP of the transpacific path is roughly \SI{1.75}{GiB} at
\SI{100}{Gbps} and a \SI{150}{ms} round-trip time, so we raised the maximum of both
\texttt{net.ipv4.tcp\_rmem} and \texttt{net.ipv4.tcp\_wmem} to \SI{1}{GiB}, as shown in
Listing~\ref{lst:sysctl}. These limits apply per socket, and since Archaea spreads a transfer
across many parallel connections, the total socket buffer aggregated over those connections covers
the BDP of the path. On the EXAScaler file system, we set the stripe count to eight, matching the
number of OSTs the appliance exports, and the stripe size to \SI{1}{MiB}. A \SI{64}{MiB} content
block is therefore split into 64 stripes of \SI{1}{MiB}, distributed across the eight OSTs. Multiple
RPCs can run in parallel to use the OSTs efficiently.

\begin{lstlisting}[caption={Kernel parameters configured on the DTNs.},
  label=lst:sysctl,float,style=paper]
net.ipv4.tcp_rmem = 4096 87380 1073741824
net.ipv4.tcp_wmem = 4096 87380 1073741824
\end{lstlisting}

\subsubsection{Workload}
For each path we transferred a \SI{1}{TB} file using the three I/O modes (buffered, direct, and
asynchronous), and observed (1) the effective end-to-end throughput reported by the Archaea client,
defined as the volume of data transferred divided by the time the transfer took, and (2) the
instantaneous throughput of the network interface reported by vnStat.

On the loopback path, we transferred once per mode. On the transpacific path, we repeated the
transfers three times per mode and report the fastest run of the ones that completed.
Since the time available on the show floor was limited, we aborted buffered I/O transfers once their
throughput had become steady. The effective throughput reported for buffered I/O therefore divides
the portion of the file that was transferred up to the abort by the time that portion took.
On the transpacific path, we additionally transferred several \SI{1}{TB} files concurrently, up to
six files for buffered I/O, four for direct I/O, and two for asynchronous I/O.

\subsection{Loopback Path}\label{sec:loopback}
\begin{figure}
  \centering
  \includegraphics{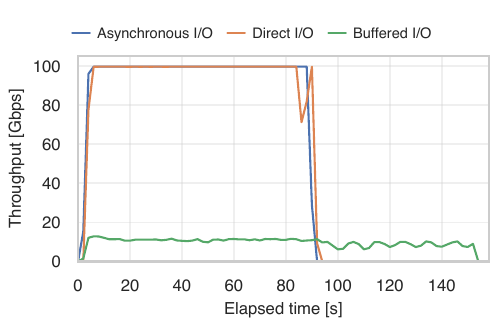}
  \caption{Instantaneous network throughput during a single \SI{1}{TB} file transfer over the
  loopback path.}
  \label{fig:loopback-single}
\end{figure}

To measure the disk-to-disk performance without the influence of the wide-area network, we first
used a loopback path closed within the SC24 venue, transferring between local NVMe SSDs at both
ends. Table~\ref{tab:loopback} reports the results, and Fig.~\ref{fig:loopback-single} shows the
instantaneous throughput of each transfer. Asynchronous I/O and direct I/O reached \SI{98.56}{Gbps}
and \SI{97.60}{Gbps}, respectively, close to the \SI{100}{Gbps} wire rate, without exhibiting an I/O
bottleneck, whereas buffered I/O reached only \SI{8.96}{Gbps}.

The write-back of the page cache accounts for this gap. Under buffered I/O, the receiver
copies the incoming data into the page cache, and the sustained throughput is therefore governed by
the rate at which the kernel flushes the dirty pages to the SSDs.

\begin{table}
  \centering
  \caption{Effective throughput of a single \SI{1}{TB} file transfer over the loopback path.}
  \label{tab:loopback}
  \begin{tabular}{lrr}
    \toprule
    I/O mode & Throughput [GB/s] & Throughput [Gbps] \\
    \midrule
    Buffered I/O & 1.12 & 8.96 \\
    Direct I/O & 12.20 & 97.60 \\
    Asynchronous I/O & 12.32 & 98.56 \\
    \bottomrule
  \end{tabular}
\end{table}

\begin{table}[t]
  \centering
  \caption{Effective throughput of a single \SI{1}{TB} file transfer over the transpacific path.}
  \label{tab:wan}
  \begin{tabular}{lrr}
    \toprule
    I/O mode & Throughput [GB/s] & Throughput [Gbps] \\
    \midrule
    Buffered I/O & 1.51 & 12.08 \\
    Direct I/O & 4.45 & 35.60 \\
    Asynchronous I/O & 11.26 & 90.08 \\
    \bottomrule
  \end{tabular}
\end{table}

\begin{figure}
  \centering
  \includegraphics{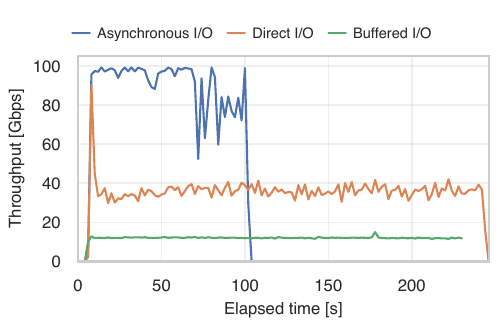}
  \caption{Instantaneous network throughput during a single \SI{1}{TB} file transfer over the
  transpacific path.}
  \label{fig:wan-single}
\end{figure}

\subsection{Transpacific Path (US--JP)}\label{sec:wan}
We then measured the transpacific path, over which the receiving DTN writes to EXAScaler.
Table~\ref{tab:wan} reports the effective throughput of a single \SI{1}{TB} file transfer.
Asynchronous I/O reached \SI{90.08}{Gbps} (\SI{11.26}{GB/s}), whereas buffered and direct I/O
reached only \SI{12.08}{Gbps} and \SI{35.60}{Gbps}, respectively. As shown in
Fig.~\ref{fig:wan-single}, the instantaneous throughput of the asynchronous transfer rises quickly
to nearly \SI{100}{Gbps}. The buffered I/O curve in the same figure breaks off while the transfer is
still in progress, because we aborted that run once its throughput had become steady at around
\SI{12}{Gbps}.

As on the loopback path, buffered I/O is limited by the write-back of the page cache. Direct I/O,
which reached the wire rate on local NVMe, attained only \SI{35.60}{Gbps} against EXAScaler. Because
parallel direct I/O is enabled on the client (\texttt{llite.*.parallel\_dio=1}, the default in our
EXAScaler configuration), even a single synchronous request distributes its RPCs across all OSTs in
parallel. The limitation is that direct I/O keeps only one request in flight, which caps the number
of RPCs concurrently outstanding at the OSTs and hence the storage throughput. This concurrency can
be increased either by using a larger I/O request size or by keeping more requests in flight.
Enlarging the request size (i.e., the content block size in Archaea) raises the number of
outstanding RPCs. Too large a block size, however, reduces the overlap of network transfer with
storage I/O and causes load imbalance across the parallel TCP connections, lowering the effective
throughput. Asynchronous I/O instead keeps many requests in flight, allowing it to raise concurrency
without growing the block size.

\begin{figure}
  \centering
  \subfloat[Buffered I/O]{\includegraphics{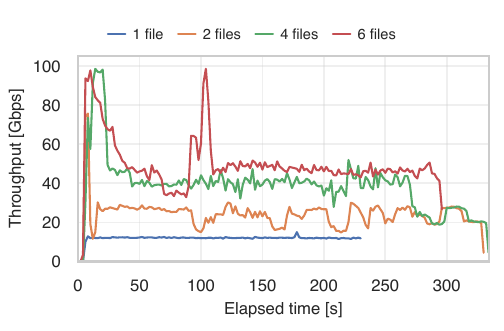}\label{fig:wan-multi-buffered}}
  \\[1ex]
  \subfloat[Direct I/O]{\includegraphics{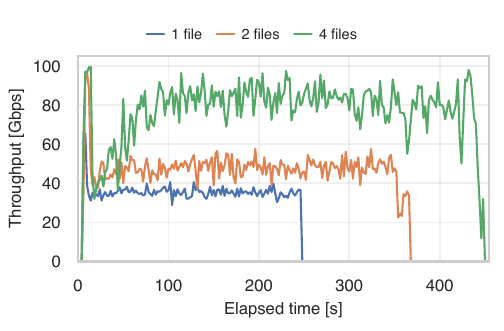}\label{fig:wan-multi-direct}}
  \\[1ex]
  \subfloat[Asynchronous I/O]{\includegraphics{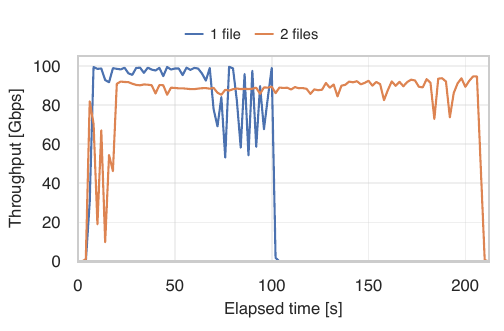}\label{fig:wan-multi-async}}
  \caption{Instantaneous network throughput during multi-file transfers over the transpacific
  path, for each I/O mode and degree of concurrency.}
  \label{fig:wan-multi}
\end{figure}

Figure~\ref{fig:wan-multi} reports the multi-file experiments, in which several \SI{1}{TB} files
were transferred concurrently. The throughput values reported below are median instantaneous
throughputs over each run.
Buffered I/O sustained \SI{12.1}{Gbps} with one file, \SI{25.3}{Gbps} with two, and \SI{39.6}{Gbps}
with four, but six files brought it only to \SI{47.0}{Gbps}. Direct I/O
sustained \SI{35.1}{Gbps}, \SI{47.6}{Gbps}, and \SI{80.7}{Gbps} with one, two, and four files.
Asynchronous I/O already sustained \SI{96.5}{Gbps} with a single file, close to the wire rate, and
\SI{88.9}{Gbps} in aggregate with two. Here too, the buffered I/O runs were aborted early, so
Fig.~\ref{fig:wan-multi}\subref{fig:wan-multi-buffered} shows only the initial portion of each
transfer.

How the three modes respond to concurrency follows from the same mechanisms as in the single-file
case. Buffered I/O roughly doubles from one file to two and continues to grow up to four, but the
shared cost of the page-cache write path limits all concurrent transfers, so each additional file
adds less throughput than the one before it and the aggregate flattens near \SI{47}{Gbps}. Direct
I/O avoids this. Since every file is striped across all OSTs, more files deepen the RPC queue of
each OST, so the aggregate grows with the file count and nears the wire rate at four. Asynchronous
I/O reaches the same per-OST depth from a single file by keeping many requests in flight, so it
saturates the OSTs with one file, and the aggregate even falls slightly with two.

Because of its large round-trip time (\SI{150}{ms}), the transpacific path is the most demanding
scenario within the scope of RED-ONION. Asynchronous I/O nevertheless moved a single file from local
NVMe to EXAScaler at \SI{90}{Gbps} on this \SI{100}{Gbps} link, which shows that the I/O
bottleneck and the single-file inefficiency have been largely resolved.

\section{Related Work}\label{sec:related}

\subsection{High-Performance Data Transfer Tools}
Many specialized tools have been developed for moving bulk data over wide-area networks.
The best known of them is GridFTP, a protocol that extends FTP with parallel TCP streams,
striping across multiple nodes, and partial-file transfers, and that remains the de facto standard
for large-scale scientific data movement. Its most widely used implementation is the one distributed
with the Globus Toolkit~\cite{Allcock2005}.
More recent designs target multicore platforms. The mdtmFTP tool
introduces multicore-aware I/O scheduling, zero-copy, and asynchronous I/O to drive \SI{100}{Gbps}
interfaces~\cite{Zhang2018}, and XRootD provides a scalable architecture for federated, concurrent
data access~\cite{Dorigo2005}. All of these tools primarily target a use case in which a dataset of
many files is moved between pools of DTNs, and the throughput reported for them is an aggregate over
concurrent file transfers rather than the speed of any individual
transfer~\cite{Fajardo2021,Kettimuthu2018}.

In practice, however, single-file transfers dominate. An analysis of Globus transfer logs found that
\SI[number-unit-product={}]{63}{\percent} of all transfers involve one file~\cite{Liu2018}. The
demand that RED-ONION targets, moving one terabyte-class file as fast as the link allows, has grown
with the output sizes of exascale simulations and of modern experimental instruments. Globus has
responded by adding
chunking~\cite{Zheng2024}, which partitions a single large file into disjoint chunks that are
carried concurrently by several pairs of DTNs. RED-ONION pursues the same goal without scaling out.
A transfer stays within a single pair of DTNs, and the throughput of one file comes from
parallelizing network transmission and storage I/O within the node.

Zettar zx\footnote{\url{https://www.zettar.com/}} represents the state of
the art in end-to-end transfer performance. In an ESnet evaluation on a \SI{100}{Gbps} testbed
looped between NERSC and StarLight at a round-trip time of about \SI{90}{ms}, zx moved \SI{1}{PB}
disk-to-disk in \SI{28}{h} \SI{45}{min} at a mean of \SI{77.25}{Gbps}, which is
\SI[number-unit-product={}]{93.44}{\percent} of the \SI{82.67}{Gbps} that the underlying storage was
separately measured to deliver~\cite{Kissel2020}.

\subsection{High-Performance Data Transfer Protocols}
A separate line of work abandons TCP altogether in order to sidestep its limitations on paths with a
high BDP or a high packet loss rate. The UDP-based data transfer protocol (UDT) implements its own
reliability and congestion control above UDP and exposes them through a socket library, so that
applications gain these benefits without kernel modifications~\cite{Gu2007}. The proprietary Aspera
fast, adaptive, and secure protocol (FASP) likewise places rate and congestion control above
UDP~\cite{Munson2011}. The high-performance and flexible protocol (HpFP) is a more recent UDP-based
design that combines precise packet pacing with
retransmission control driven by continuous monitoring of loss and
latency~\cite{Murata2016a,Murata2016b}. Yet another line of work bypasses the kernel network stack
entirely, using zero-copy remote direct memory access (RDMA) over Converged Ethernet or InfiniBand
to drive wide-area transfers at near-wire rate while minimizing host CPU
involvement~\cite{Kissel2012}. A survey of efficient transfer protocols for high-bandwidth research
networks likewise reports that RDMA-based transport can saturate a \SI{40}{Gbps} path at low CPU
cost~\cite{Tierney2012}.

These designs target networks with high latency, a high packet loss rate, or both. On the
dedicated network that RED-ONION assumes, packet loss is essentially zero, and even at a high
round-trip time TCP can fill the available bandwidth once the socket buffers and the congestion
control algorithm are tuned. The Linux network stack is also under continuous improvement. With
zero-copy transmission and packet pacing on a recent kernel, a single TCP connection has been
reported to sustain \SIrange{40}{50}{Gbps} on wide-area paths of up to \SI{104}{ms} round-trip time,
essentially independently of the round-trip time~\cite{Schwarz2024}. We therefore judged that a
specialized protocol was unnecessary, and RED-ONION transfers data over TCP on the kernel network
stack. This choice also spares operators the deployment and maintenance cost of a custom protocol
stack on every DTN.

\subsection{Data Transfer Nodes and the Science DMZ}
At the architectural level, the \emph{Science DMZ} design pattern places dedicated,
performance-tuned DTNs at a network edge that is free of general-purpose
security and policy bottlenecks, complemented by continuous performance
monitoring~\cite{Dart2013}. Such monitoring spans multiple administrative domains, for which the
perfSONAR service-oriented framework has become the standard tool for diagnosing end-to-end path
problems~\cite{Hanemann2005}. RED-ONION is an implementation of this design pattern,
with DTNs at both ends of a dedicated wide-area network that bypasses the general-purpose security
and policy devices of the campus network.

\section{Conclusions and Future Work}\label{sec:conclusion}
We presented RED-ONION, a high-speed disk-to-disk data transfer system that connects
research facilities, on campus and beyond, to a computing center over a dedicated high-bandwidth
network, and we described
the end-to-end performance optimizations, spanning the data transfer software, the operating system,
and the storage system, required to achieve wire-rate single-file transfer. In an experiment on a
prototype deployment over a transpacific path between Atlanta and Tokyo with a \SI{150}{ms}
round-trip time, asynchronous I/O transferred a single \SI{1}{TB} file disk-to-disk at
\SI{90.08}{Gbps}, whereas buffered and direct I/O reached only \SI{12.08}{Gbps} and
\SI{35.60}{Gbps}, respectively. Buffered I/O is limited by the page-cache write path, and direct I/O
keeps too few requests in flight to saturate the OSTs. Asynchronous I/O sustains enough concurrent
RPCs to approach the wire rate from a single file, without enlarging the request size or aggregating
throughput over many files.

In future work, we will look beyond transfer performance to what the system needs in production.
This article focused on achieving and evaluating data transfer performance between sites, but
several problems remain before the system can be put into operation. An operational deployment
serves many users transferring many datasets at the same time, so data transfer requests must be
queued and scheduled to keep them from contending for the network between sites. As the system is
extended to more sites, it must also become multi-tenant, which calls for a design of how the DTNs,
the network, and the storage are shared across sites and across users.

\section*{Acknowledgments}
This work was partly supported by JST ACT-X Grant Number JPMJAX24M6 and JSPS KAKENHI Grant Numbers
JP23K16890 and JP21K11912. The performance measurements reported in this paper were conducted at SC24
as SCinet Network Research Exhibition demonstration SC24-NRE-037. The authors would like to thank
SCinet and its volunteers for providing show-floor connectivity, and the National Institute of
Information and Communications Technology (NICT) for coordinating the international path.
In this work, we used the ``mdx II: a platform for building data-empowered society''.

\bibliographystyle{IEEEtran}
\bibliography{references}

\clearpage

\begin{IEEEbiography}[{\includegraphics[width=1in,height=1.25in,clip,keepaspectratio]{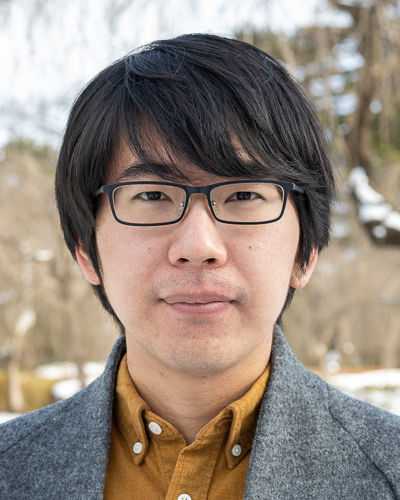}}]{Keichi Takahashi}
(Member, IEEE) received the B.E., M.S., and Ph.D. degrees from the University of Osaka, Osaka,
Japan, in 2014, 2016, and 2019, respectively. In 2018, he was a Visiting Scholar with Oak Ridge
National Laboratory, Oak Ridge, TN, USA. From 2019 to 2021, he was an Assistant Professor with the
Nara Institute of Science and Technology, Nara, Japan, and from 2021 to 2024, with Tohoku
University, Sendai, Japan. He is currently an Associate Professor with the University of Osaka,
Osaka, Japan. His research interests include high-performance computing and parallel and
distributed computing.
\end{IEEEbiography}

\begin{IEEEbiography}[{\includegraphics[width=1in,height=1.25in,clip,keepaspectratio]{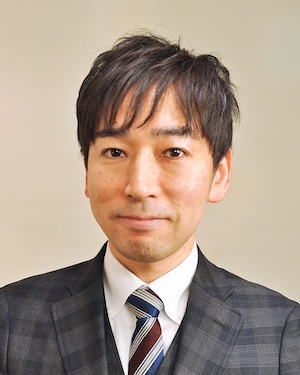}}]{Hiroaki Kataoka}
received the B.E. and M.E. degrees in information science and technology from Waseda University,
Tokyo, Japan, in 2003 and 2005, respectively. Since 2005, he has been with NEC Corporation, where he
has been engaged as a systems engineer in system design and deployment. His work focuses on ICT
infrastructure for scientific research.
\end{IEEEbiography}

\begin{IEEEbiography}[{\includegraphics[width=1in,height=1.25in,clip,keepaspectratio]{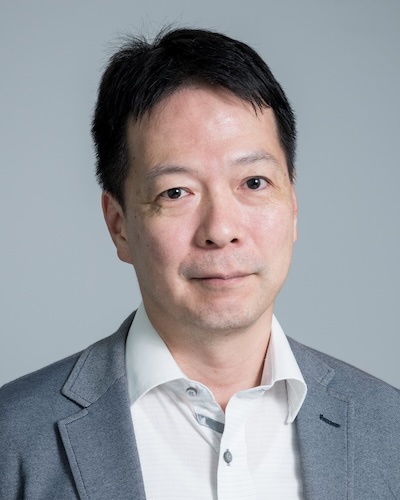}}]{Takeo Hosomi}
received the B.E. and M.E. degrees from Kyoto University, in 1992 and 1994,
respectively, and the Ph.D. degree in information science and technology from the University of
Osaka, in 2022. Since 2018, he has been a Senior Principal Researcher with NEC Corporation. His
research interests include computer architecture and high-performance computing.
\end{IEEEbiography}

\begin{IEEEbiographynophoto}{Ayahiro Takaki}
received the B.S. degree from Kyushu University, Fukuoka, Japan, in 2002, and the M.E. degree from
the Nara Institute of Science and Technology, Nara, Japan, in 2004. He withdrew from the doctoral
program at the Nara Institute of Science and Technology, co-founded CLEALINK TECHNOLOGY Co., Ltd.,
in 2005, and has been with the company for over 20 years. His research and development interests
include high-performance communication software and security.
\end{IEEEbiographynophoto}

\begin{IEEEbiographynophoto}{Yasunori Kakizawa}
received the B.E. degree from the National Institute of Technology, Kisarazu College, Kisarazu,
Japan, in 2007, and the M.S. degree from the Japan Advanced Institute of Science and Technology,
Ishikawa, Japan, in 2009. From 2008 to 2011, he was a Research Assistant with the National
Institute of Information and Communications Technology, Kyoto, Japan. Since 2011, he has been with
CLEALINK TECHNOLOGY Co., Ltd., Kyoto, Japan. His research interests include high-speed networking
technologies and natural language processing.
\end{IEEEbiographynophoto}

\begin{IEEEbiography}[{\includegraphics[width=1in,height=1.25in,clip,keepaspectratio]{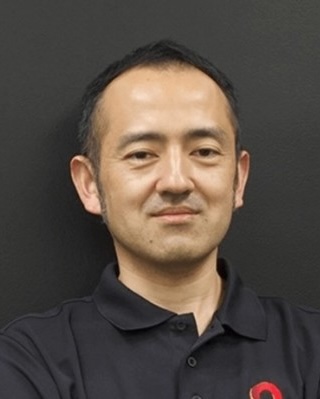}}]{Shuichi Ihara}
received the M.S. degree in Information, Electrical and Electronic Engineering from
Kumamoto University, Kumamoto, Japan, in 2001. In the same year, he joined Sun Microsystems,
Inc. In 2010, he joined DataDirect Networks Japan. He is currently a Principal Engineer with
DDN, working on high-performance computing and storage systems, with a focus on I/O performance
and optimization. He has contributed to the development and performance evaluation of Lustre and
large-scale storage systems through research publications and industry projects. His research
and engineering interests include high-performance computing, parallel file systems,
high-performance I/O, storage performance optimization, and emerging AI workloads.
\end{IEEEbiography}

\begin{IEEEbiography}[{\includegraphics[width=1in,height=1.25in,clip,keepaspectratio]{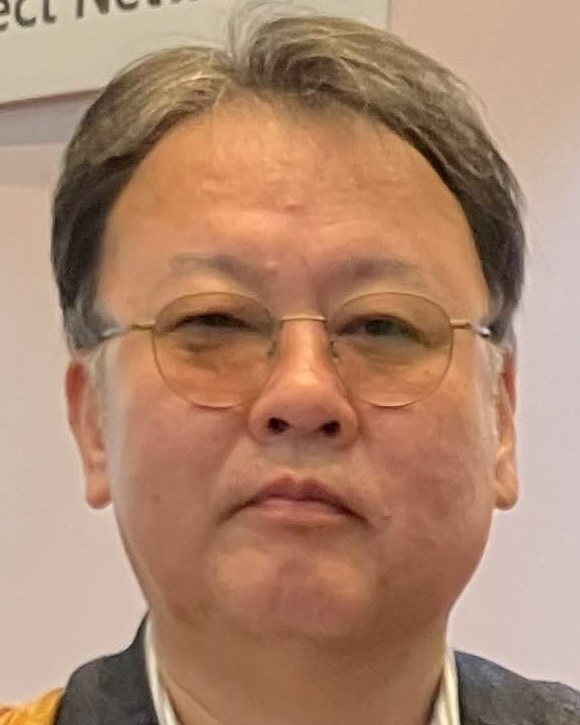}}]{Nobuaki Hashizume}
received the bachelor's degree in economics from the School of Political Science and Economics,
Meiji University, Tokyo, Japan, in 1990. From 1990 to 1993, he was a System Engineer with
Matsushita Computer Systems. In 1993, he joined Sun Microsystems, where he ultimately served as a
Principal Field Technologist for high-performance and cloud computing. From 2010 to 2011, he was a
Senior HPC Specialist with Dell. In 2011, he joined DataDirect Networks Japan, where he is
currently a Presales Engineer specializing in storage systems for high-performance computing and
artificial intelligence.
\end{IEEEbiography}

\begin{IEEEbiography}[{\includegraphics[width=1in,height=1.25in,clip,keepaspectratio]{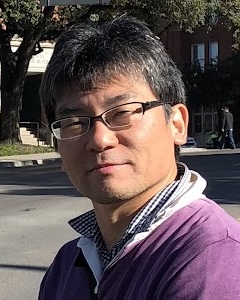}}]{Susumu Date}
(Member, IEEE) received the B.E., M.E., and Ph.D. degrees from the University of Osaka, in 1997,
2000, and 2002, respectively. He was an Assistant Professor with the Graduate School of
Information Science and Technology, the University of Osaka, from 2002 to 2005. He was a
Visiting Scholar with the University of California at San Diego, in 2005. From 2005 to 2008, he
was a Specially-Appointed Associate Professor in the internationalization of education with the
Graduate School of Information Science and Technology, the University of Osaka. From 2008 to
2023, he was an Associate Professor with the Cybermedia Center, the University of Osaka. From
2023 to 2024, he was a Professor with the Cybermedia Center. Since 2024, he has been a Professor
with the D3 Center, the University of Osaka. His research field is computer science. His current
research interests include cloud, cluster, grid, high-performance computing, and their
applications. He is a member of IPSJ.
\end{IEEEbiography}

\EOD

\end{document}